\documentclass[conference]{IEEEtran}
\IEEEoverridecommandlockouts
\usepackage{booktabs}
\usepackage{cite}
\usepackage{amsmath,amssymb,amsfonts}
\usepackage{algorithmic}
\usepackage{graphicx}
\usepackage{textcomp}
\usepackage{xcolor}
\usepackage{tabularx} 
\usepackage{makecell}

\def\BibTeX{{\rm B\kern-.05em{\sc i\kern-.025em b}\kern-.08em
    T\kern-.1667em\lower.7ex\hbox{E}\kern-.125emX}}
\begin{document}

\title{Confidence-Based Self-Training for EMG-to-Speech: Leveraging Synthetic EMG for Robust Modeling}

\author{
\IEEEauthorblockN{
Xiaodan Chen\IEEEauthorrefmark{1}\IEEEauthorrefmark{2}\IEEEauthorrefmark{3}, 
Xiaoxue Gao\IEEEauthorrefmark{2}, 
Mathias Quoy\IEEEauthorrefmark{1}\IEEEauthorrefmark{3}, 
Alexandre Pitti\IEEEauthorrefmark{1}\IEEEauthorrefmark{3}, 
Nancy~F.~Chen\IEEEauthorrefmark{2}\IEEEauthorrefmark{3}
}
\IEEEauthorblockA{\IEEEauthorrefmark{1}ETIS, CY Cergy Paris Université - ENSEA – CNRS, UMR 8051, 2 Av. Adolphe Chauvin, 95300 Pontoise, France}
\IEEEauthorblockA{\IEEEauthorrefmark{2}Institute for Infocomm Research, A*STAR, 1 Fusionopolis Way, \#20-10 Connexis North Tower, Singapore 138632}
\IEEEauthorblockA{\IEEEauthorrefmark{3}IPAL (International Research Laboratory on Artificial Intelligence), CNRS, Singapore}
\IEEEauthorblockA{
Email: xiaodan.chen@cyu.fr, gao\_xiaoxue@i2r.a-star.edu.sg, mathias.quoy@cyu.fr, \\
alexandre.pitti@cyu.fr, nfychen@i2r.a-star.edu.sg
}
\vspace{-1cm}
\thanks{Xiaoxue Gao is the corresponding author.}
}

\maketitle

\begin{abstract}
Voiced Electromyography(EMG)-to-Speech (V-ETS) models reconstruct speech from muscle activity signals, facilitating applications such as neurolaryngologic diagnostics.
Despite its potential, the advancement of V-ETS is hindered by a scarcity of paired EMG-speech data. To address this, we propose a novel Confidence-based Multi-Speaker Self-training (CoM2S) approach, along with a newly curated Libri-EMG dataset. This approach leverages synthetic EMG data generated by a pre-trained model, followed by a proposed filtering mechanism based on phoneme-level confidence to enhance the V-ETS model through the proposed self-training techniques. Experiments demonstrate our method improves phoneme accuracy, reduces phonological confusion, and lowers word error rate, confirming the effectiveness of our CoM2S approach for V-ETS. 
In support of future research, we will release the codes and the proposed Libri-EMG dataset—an open-access, time-aligned, multi-speaker voiced EMG and speech recordings.
\end{abstract}


\section{Introduction}
Voiced Electromyography-to-Speech (V-ETS) aims to reconstruct speech from muscle activity signals, facilitating the interpretability and controllability by simultaneously capturing articulatory muscle signals and speech \cite{lee2025articulatoryfeaturepredictionsurface}.
V-ETS greatly supports AI in healthcare applications such as neurolaryngologic diagnostics \cite{Silva_Moraes_Pernambuco_Moraes_Balata_2014}, research endeavors in manipulable speech generation \cite{Gao_Birkholz_Li_2024, Chung_Kang_2024} and articulatory-to-acoustic decoding \cite{Kearney_Guenther_2019, Guenther_1995}.
Unlike silent ETS, which maps unvoiced muscle activity to silent speech and lacks direct exposure to audible speech, V-ETS captures a stronger relationship between muscle signals and actual speech. With audible speech informed, V-ETS enables more precise modeling of speech-related neuromuscular activity. While silent ETS may benefit from speech-aligned cues informed by V-ETS, the latter offers complementary advantages for studying speech production and phoneme articulation. Therefore, V-ETS is essential for advancing our understanding of the physiological basis of speech production and phoneme articulation.


Although research on Electromyography-to-Speech (ETS) has been growing \cite{lee2025articulatoryfeaturepredictionsurface, gaddy-klein-2020-digital, Gaddy_EECS-2022-68, Toth_Wand_Schultz_2009, gaddy2021improvedmodelvoicingsilent, ren2024diffets, ren2023selflearning, Scheck2023SUE2S, wu2024emgtospeechnecklaceformfactor, 10781707, 10363027, Diener_Vishkasougheh_Schultz_2020, 8114359, 8578038, 7280404, scheck23_interspeech, Wand_Janke_Schultz_2014, Schultz_Wand_2010}, the availability of EMG-speech datasets remains limited due to labor-intensive and costly data collection from human participants. Currently, open-access voicing EMG-speech datasets include only 2 hours of recordings from \cite{Wand_Janke_Schultz_2014}, 20 hours from \cite{gaddy-klein-2020-digital}, and 9.5 hours from \cite{Diener_Vishkasougheh_Schultz_2020}. More critically, differences in signal recording configurations across datasets make them incompatible with one another. For instance, \cite{Diener_Vishkasougheh_Schultz_2020} used a high-density setup with 40 electrodes, while \cite{gaddy-klein-2020-digital} employed only 8, and \cite{Wand_Janke_Schultz_2014} recorded with just 6. These inconsistencies introduce signal mismatches. Such data scarcity and incompatibility poses significant challenges, particularly in machine learning and deep learning modeling which may require consistent and meaningful EMG inputs.

To address this, some studies have explored data augmentation techniques to improve model performance. For instance, \cite{ren2023selflearning} investigated self-learning and active-learning strategies to expand datasets, demonstrating that human-in-the-loop corrections significantly enhanced model performance. \cite{Scheck2023SUE2S} introduced SU-ETS, a model that predicts speech units (SUs) \cite{van_Niekerk_2022} from EMG signals for speaker-independent synthesis by incorporating a voice conversion model. However, these augmentation approaches primarily rely on reusing existing speech content without introducing new phonological knowledge (e.g., transforming the same sentence into different voices without changing its wording \cite{Scheck2023SUE2S} or filtering the original dataset before retraining \cite{ren2023selflearning}—and often still require additional human effort, such as re-recording speech to improve data quality \cite{ren2023selflearning}.
Moreover, existing ETS models have yet to surpass the best published voiced WER of 23.3\% \cite{Gaddy_EECS-2022-68}, highlighting the need for alternative methods to increase dataset size and new training strategies.

A promising approach to address data limitation is self-training, a semi-supervised learning technique that has proven effective in other fields \cite{Kahn_Lee_Hannun_2020, Xu_Baevski_Likhomanenko_Tomasello_Conneau_Collobert_Synnaeve_Auli_2020, Gao_Yue_Li_2022}, where self-training fundamentally involves using the model’s own predictions (or outputs) to creating additional training data (pseudo-labels). However, despite its success in other tasks, self-training remains largely unexplored in the ETS domain, leaving a significant gap in research that this work seeks to address.

To bridge this gap, we propose a novel data augmentation approach and training strategy that combines self-training to extract high-quality EMG-speech time-aligned data from a large repository of open-access speech resources, LibriSpeech \cite{Librispeech}. Our method, \textbf{Co}nfidence-based \textbf{M}ulti-\textbf{S}peaker \textbf{S}elf-training
(\textbf{CoM2S}) for V-ETS, leverages a pre-trained generator model \cite{scheck23_interspeech} to generate EMG features aligned with multi-speaker speech and employs a confidence-based self-training strategy to filter high-quality synthetic samples. 
This approach effectively expands the available data without requiring additional costly EMG recordings while mitigating data mismatch issues by including session embeddings.

The main contributions of this work are as follows:
\begin{itemize}
    \item Threshold-Tuned Self-Training: We incorporate self-training into the V-ETS domain by systematically evaluating phoneme accuracy thresholds to optimize the quality-quantity tradeoff of synthetic EMG data;
    \item Open-Source Dataset: We introduce Libri-EMG, an 8.3-hour open-access high quality multi-speaker voicing dataset, expanding multi-speaker EMG data to support further research in EMG-based speech modeling \cite{chen2025vets_github};
    \item Extensive Experiments: We analyze the impact of different training strategies and data ratios, demonstrating that a 1:1 real-to-synthetic mixing achieves optimal performance and outperforms the best published \cite{Gaddy_EECS-2022-68} voiced WER of 23.3\%.
\end{itemize}

\begin{figure*}[th!]
  \centering
  \includegraphics[width=0.8\linewidth]{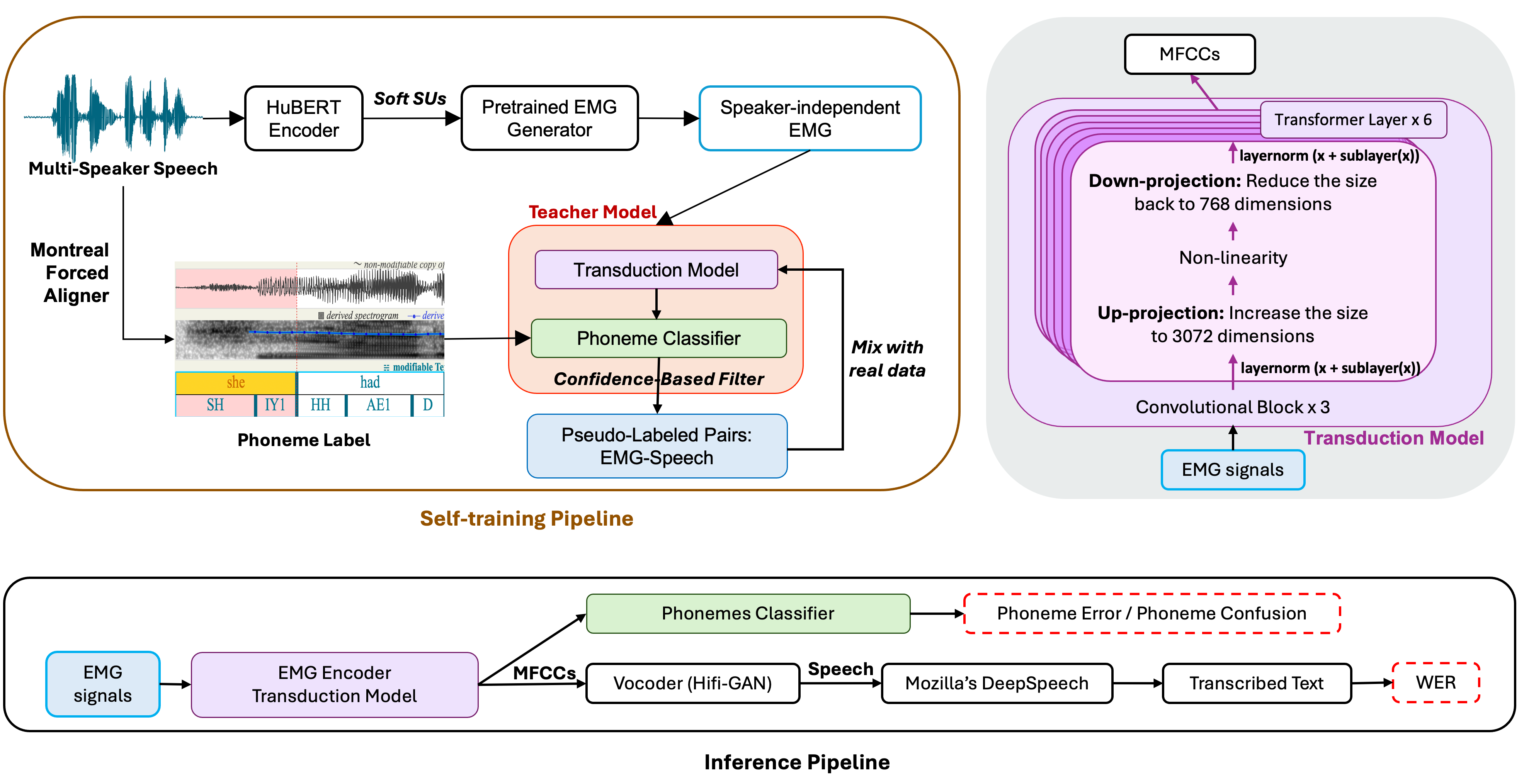}
  \caption{Top left: Overview of our CoM2S approach for V-ETS. We employ a GAN-based EMG generator \cite{scheck23_interspeech} conditioned on speaker-independent Soft Speech Units generated from HuBERT Encoder \cite{van_Niekerk_2022}, with learnable session embeddings accounting for variations in electrode configurations. The generated EMG data then undergoes preprocessing, including upsampling and inverse transformation, to align with real EMG signals as described in Sec. \ref{generation_libriEMG}. A pretrained transduction model together with the pretrained classifier \cite{gaddy2021improvedmodelvoicingsilent} serves as the teacher model, filtering synthetic samples based on phoneme accuracy. Only high-confidence synthetic data is retained and proportionally mixed with real EMG data for self-training, ensuring robust adaptation while maintaining phonetic consistency. Top right: baseline transduction model architecture \cite{gaddy-klein-2020-digital, gaddy2021improvedmodelvoicingsilent}. Bottom: inference pipeline.}
  \label{fig:Overview}
\end{figure*}

\section{Methods}

Given the scarcity of EMG-speech time-aligned data, in this work, we investigate whether synthetic data—paired with confidence-based filtering—can be used effectively in a self-training approach to V-ETS modeling.

Fig. \ref{fig:Overview} illustrates the CoM2S approach pipeline (top left), baseline architecture (top right), and CoM2S V-ETS inference pipeline (bottom). Our CoM2S approach begins by generating time-aligned EMG features from speech, simulating the EMG modality across diverse speakers. We then employ a self-training approach in which the baseline model together with a confidence-based filter generates pseudo-labels for synthetic inputs. We further explore how filter thresholds affect model performance, and whether small but high-confidence subsets may outperform larger, less filtered ones. Additionally, we investigate the impact of mixing filtered synthetic data with real EMG data in various proportions during self-training. Finally, we propose a train-from-scratch approach using real paired EMG-speech inputs and synthetic ones to evaluate synthetic multi-speaker EMG-speech data quality that jointly processes real and synthetic paired EMG-speech data, enabling direct comparison of their contributions to model learning.


We aim to identify optimal strategies for leveraging synthetic data and confidence filtering to enhance V-ETS model in low-resource scenarios. 


\subsection{Generation of Time-aligned EMG Features from Multi-Speaker Speech}\label{generation_libriEMG}

We adopted the pretrained generator from a generator model \cite{scheck23_interspeech}, which builds upon \cite{Morrison_Kumar_Kumar_Seetharaman_Courville_Bengio_2021}. The EMG generator is conditioned on speech content representations extracted by voice conversion (VC) models, allowing it to take speaker-independent Soft Speech Units (Soft SUs) as input \cite{van_Niekerk_2022}\cite{Morrison_Kumar_Kumar_Seetharaman_Courville_Bengio_2021}. While the generator does not differentiate between speakers at the voice level, it incorporates learnable session embeddings to compensate for variations in electrode configurations across different EMG recording sessions. Since our CoM2S approach is not constrained to a specific recording parameter, we assign session indices randomly but evenly to match the distribution used in our baseline ETS model.
By conditioning the generator on speaker-independent Soft SUs \cite{Scheck2023SUE2S}, we maintain speaker invariance at the speech content level while leveraging learnable session embeddings to address variability in electrode configurations, aligning with that of our baseline ETS model.

\subsection{Self-training Pseudo-Labeling with Confidence-Based Filtering}

We propose to utilize the pretrained generator \cite{scheck23_interspeech} to generate speaker-independent synthetic EMG features conditioned on multi-speaker speech. 
However, these generated EMG features are not guaranteed to be speaker-consistent or fully faithful to natural EMG patterns. As such, although the accompanying speech is real and labeled, the pair (synthetic EMG-real speech) is not ground-truth-aligned in the conventional sense. Therefore, we treat the real speech (or its derived features such as MFCCs or phoneme labels) as pseudo labels for the synthetic EMG input. By applying confidence-based filtering, we select only the synthetic EMG–speech pairs for which the transduction model produces confident outputs. This filtering acts as a form of pseudo-label validation, ensuring that only plausible synthetic EMG inputs with reliable label alignment are used for subsequent self-training.

In our proposed CoM2S approach, confidence is measured by the phoneme accuracy of generated samples, and only synthetic data with a phoneme error below a predefined threshold is retained for self-training. 
As a result, on top of the output of the transduction model, we used a pretrained phoneme classifier \cite{gaddy-klein-2020-digital}\cite{Scheck2023SUE2S} as shown on the top left in Fig. \ref{fig:Overview}.  
The phoneme error calculation is based on cross-entropy loss between predicted phoneme probabilities and the target phoneme sequence\cite{gaddy2021improvedmodelvoicingsilent, scheck23_interspeech}:

\begin{equation}
    L_{\text{phoneme}} = -\sum_{t=1}^{T} \sum_{c=1}^{C} y_{t,c} \log(p_{t,c})
    \label{eq:phoneme_loss}
\end{equation}
where $T$ represents sequence length (a.k.a number of phoneme steps), $C$ number of phoneme classes (ARPABet phonemes \cite{mfa_english_us_arpa_acoustic_2024}), $y_{t,c}$ ground truth one-hot encoded phoneme at step $t$ (1 for correct class, 0 otherwise), $p_{t,c}$ the predicted probability of phoneme class $c$ at step $t$. The pretrained model serves as a teacher model, guiding the selection of high-quality synthetic data generated by the EMG generator. 

\subsection{Phoneme Loss Threshold Exploration for Training Data Filtering}
\label{section: threshold}

When working with synthetic EMG-speech pairs, data quality can vary significantly depending on how well the generated speech matches intended phonemic content. Low-quality synthetic data may introduce noise during training, hindering model generalization. To address this, we explored whether filtering synthetic samples based on phoneme prediction loss could serve as an effective proxy for confidence. The core motivation was to assess whether prioritizing high-confidence examples could improve training efficiency and model performance, even at the cost of reducing the overall training data volume. This approach aims to balance the trade-off between data quantity and quality in scenarios where large-scale high-fidelity synthetic data is difficult to guarantee. Therefore, we investigate the impact of different phoneme loss thresholds on model performance. 



Using the synthetic data generation approach described in Section~\ref{generation_libriEMG}, we created training subsets by applying different phoneme loss thresholds, which were then used to train separate instances of the baseline ETS model. Performance was evaluated across various test sets to assess the relationship between training data confidence and downstream performance.
Again, the goal was to determine the optimal phoneme loss threshold that balances data quality and volume, thereby improving training efficacy and downstream V-ETS performance.

\subsection{Mix-proportion Exploration for Self-Training}
\label{section:Mix-proportion}

To further improve training efficiency, we proportionally mix the filtered synthetic data with real EMG data, ensuring a balanced representation of both real and synthetic signals. The inclusion of real baseline EMG-speech data is crucial, as our self-training method relies on both real and synthetic sources to strike a balance between authentic V-ETS mappings and data diversity from multi-speaker synthetic samples. By integrating both, we aim to leverage the robustness of real data while enhancing generalization with synthetic data, ensuring that the model remains grounded in real EMG patterns while benefiting from additional training examples. By carefully tuning the real-to-synthetic ratio, we aim to optimize the trade-off between model generalization and training stability. The overall pipeline of our approach is illustrated in Fig. \ref{fig:Overview}. After supporting evidence, we investigate the relationship between dataset scale and self-training efficacy by progressively increasing the total training volume while maintaining the established ratio.



\subsection{Train-from-Scratch Approach for Synthetic Data Investigation}
\label{trainfromscratch}
In our previous investigation, we did not use purely real data for inference purposes because the pretrained model from \cite{gaddy2021improvedmodelvoicingsilent} had already been trained on all available real voiced data. As a result, evaluating on a fully real test set would not provide meaningful insights. Additionally, to rigorously assess the contribution of synthetic EMG-speech data, we need a model where synthetic data is fully integrated into the learning process rather than used as a secondary refinement step.

To address both concerns, we propose a train-from-scratch method: using a mix of real and synthetic data based on the best mix ratio. This ensures that the voiced test dataset remains entirely unseen during training while also allowing us to directly compare the impact of synthetic data on V-ETS conversion. Unlike self-training, where synthetic data is introduced after pretraining, this approach ensures that both real and synthetic data contribute equally to the learning process from the beginning. For a fair comparison, we compared with the baseline model trained exclusively on real voiced EMG-speech data, as described in \cite{gaddy2021improvedmodelvoicingsilent}. Both models share the same architecture and training procedures to ensure consistency. By comparing WER, we determine whether synthetic data enhances model performance beyond what can be achieved with real data alone.

\section{Experimental Setup}
\subsection{Baseline Model and Dataset}
\label{Baseline_model_data}
We used the transduction model \cite{gaddy2021improvedmodelvoicingsilent} as our baseline ETS model, a widely recognized baseline for ETS models. Additionally, to our knowledge, it is the only one that has trained and evaluated purely on voiced EMG, matching our setting. This model is originally designed for EMG recorded with eight electrodes, in line with the synthetic EMG data. As drawn on top right in Fig. \ref{fig:Overview}, its transformer-based \cite{vaswani2017attention} architecture consists of three convolutional blocks followed by six transformer layers, directly processing EMG signals as well as session embedding to predict Mel-Frequency Cepstral Coefficients (MFCCs) as output.

For the real dataset, we selected the parallel and non-parallel voiced EMG-speech data from \cite{gaddy-klein-2020-digital}, excluding the silent data, as our focus is on V-ETS conversion.

\subsection{Automatic Evaluation Matrics}




To ensure a fair and controlled comparison with the baseline ETS model \cite{Gaddy_EECS-2022-68}, we adopt an automatic evaluation method that isolates the impact of changes in the ETS transduction model. Specifically, both the phoneme classifier \cite{gaddy2021improvedmodelvoicingsilent} and the HiFi-GAN vocoder \cite{kong2020hifigangenerativeadversarialnetworks} are kept identical to those used in the baseline and are frozen during all self-training or train-from-scratch experiments, ensuring that they do not adapt to any artifacts or speaker variability introduced by synthetic inputs. This design choice guarantees that any observed improvements or degradations in performance are attributable solely to the V-ETS model and not to adaptation in downstream components.

We use three automatic metrics for evaluation: phoneme accuracy, phoneme confusion, and word error rate (WER). Phoneme accuracy and confusion are computed using the frozen phoneme classifier applied to the MFCC output of the V-ETS model and are defined as follows:

\begin{align}
  \text{confusion}(p_1, p_2) &= \frac{e_{p_1,p_2} + e_{p_2,p_1}}{f_{p_1} + f_{p_2}}
  \label{equation:Confusion}
\end{align}
\vspace{-0.4cm}
\begin{align}
  \text{accuracy}(p_1, p_2) &= \frac{e_{p_1,p_1} + e_{p_2,p_2}}{f_{p_1} + f_{p_2}}
  \label{equation:Confusion}
\end{align}

where $e_{p_1,p_2}$  denotes the number of times phoneme $p_2$ was predicted when the ground truth was $p_1$, and $f_{p_1}$ is the total number of occurrences of phoneme $p_1$ in the dataset. The WER is calculated using Mozilla’s DeepSpeech \cite{DeepSpeech} applied to the final speech waveform produced by the frozen HiFi-GAN vocoder. This pipeline mirrors the original evaluation setup in \cite{Gaddy_EECS-2022-68}, enabling direct comparison.

By keeping the classifier and vocoder fixed and trained solely on real data, we avoid introducing evaluation bias, particularly when testing models trained with synthetic data. This setup allows us to interpret changes in WER and phoneme metrics as genuine improvements in V-ETS transduction quality, not artifacts of downstream model adaptation.

\subsection{Preprocessing Generated EMG for Speech Synthesis}

To preprocess the generated EMG data before feeding to the transduction model, we apply an upsampling and reverse transformation procedure. First, we upsample the signal from its original sampling rate to the target rate using linear interpolation. This ensures temporal alignment with other EMG recordings at a unified frequency. Next, we apply reverse processing to restore the EMG signal to its original range. Since the GAN-generated EMG values are transformed via a tanh function during training \cite{scheck23_interspeech}, we apply an inverse tanh (arctanh) transformation to recover the original distribution. To avoid numerical singularities at extreme values (-1 and 1), we first clip the signal within the range $[-1+10^{-10}, -1-10^{-10}]$. The recovered EMG values are then scaled by a factor of 100, matching the amplitude distribution of real EMG data. This processing ensures that the generated EMG signals are comparable to the original recordings while maintaining the proper frequency characteristics.

\section{Results and Discussions}
\subsection{Phoneme-Error-Based Filtered Synthetic Libri-EMG Data}
\label{filter_threshold_experiment}

Using the approach described in Sec. \ref{generation_libriEMG}, we generated three subsets of synthetic EMG-speech pairs by filtering the data using phoneme loss thresholds: no filtering (Raw), loss $<$ 0.8, and loss $<$ 0.5. Each subset was then used to continue train separate baseline models, and performance was evaluated across all test sets using WER. we then trained three versions of the baseline ETS model on synthetic training subsets using the dev-clean dataset from LibriSpeech \cite{Librispeech} filtered at different phoneme loss thresholds:
\begin{table}[h]
    \centering
    \caption{Filtered Synthetic Dataset Size under Different Confidence Thresholds}
    \begin{tabular}{ll}
        \hline
        \textbf{Condition} & \textbf{Filtered Dataset Size} \\
        \hline
        Raw (no filtering) & $\sim$5.4 hours \\
        Phoneme Loss$<$0.8 & $\sim$5.0 hours \\
        Phoneme Loss$<$0.5 & $\sim$0.5 hours \\
        \hline
    \end{tabular}
    \label{tab:phoneme_loss_filtering}
\vspace{-0.3cm}
\end{table}

\begin{figure}[h]
  \centering
  \includegraphics[width=0.7\linewidth]{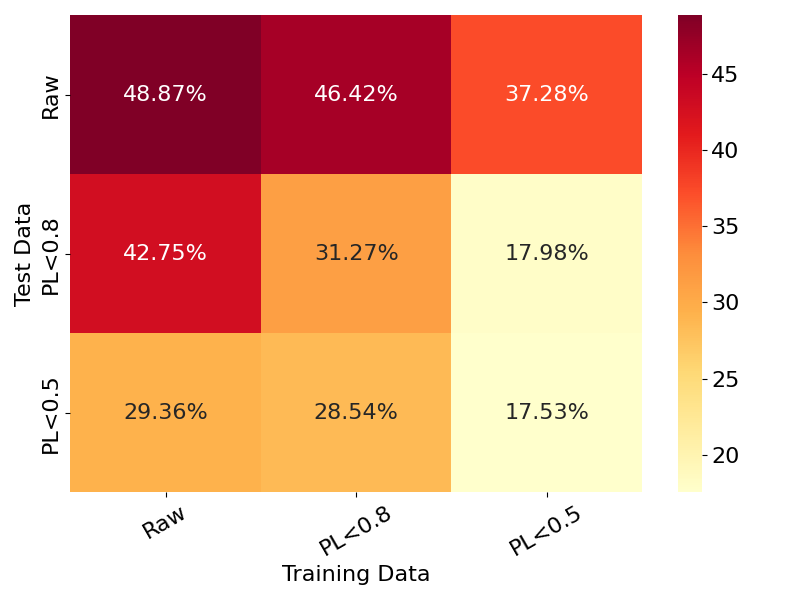}
  \caption{Performance comparison of EMG-based speech recognition models trained on different filtered subsets of self-generated data (5.4h dev-clean in LibriSpeech \cite{Librispeech}) and evaluated on corresponding test sets. Values and colors represent word error rates (WER) (lower/lighter is better). 
  }
  \label{fig:heatmap}
\vspace{-0.5cm}
\end{figure}

The evaluation metric (lower/lighter is better) in Fig. \ref{fig:heatmap} indicates that the model trained on the smallest but highest-confidence filtered subset (PL$<$0.5, $\sim$0.5h) consistently achieves the best or comparable performance across all test sets, including the full raw test set ($\sim$ 5.4h) and the filtered subsets.

Specifically, on the raw test data, the PL$<$0.5 trained model attains a WER of 29.36\%, outperforming the models trained on larger but less filtered datasets (42.75\% for PL$<$0.8 and 0.48.87\% for raw data). This suggests that training on high-confidence, filtered data enables the model to generalize better, despite the smaller training size. Similarly, on the filtered test sets (PL$<$0.8 and PL$<$0.5), the PL$<$0.5 trained model matches or slightly improves upon the performance of models trained on larger datasets, with WER of 28.54\% and 17.53\% respectively, reinforcing the benefit of data quality over quantity.

In summary, these results demonstrate that filtering training data by confidence (using phoneme loss thresholds) effectively improves model generalization and performance, even when reducing training data volume significantly.

\begin{table}[h!]
  \caption{Overview of Real and Synthetic Voiced Datasets}
  \label{tab:data}
  \centering
  \begin{tabular}{p{1cm} p{1cm} p{2cm} p{2.2cm}}
    \toprule
    \textbf{Data}      & \textbf{Speaker Number}    & \textbf{Gender} & \textbf{Dataset Size (Utterance Number)}            \\
    \midrule
    \cite{Gaddy_EECS-2022-68}                    & $1$ & Male & $7065 (\approx15.2h)$                                       \\
    Ours                    & 1532 & Male \& Female & $3514 (\approx8.3h)$                                 \\
    \bottomrule
  \end{tabular}
\vspace{-0.2cm}
\end{table}

Applying the optimal phoneme-error threshold of $<0.5$, we generated 8.3 hours of EMG-speech data, covering a diverse set of 1,532 speakers across both male and female categories \cite{chen2025vets_dataset}. This multi-speaker dataset ensures robust modeling and allows us to evaluate the generator’s ability to generalize across different speakers and recording conditions. For better visualization of the dataset used in the following experiments, we list both the real baseline data discussed in Sec. \ref{Baseline_model_data} and the synthetic data in Table \ref{tab:data}.

\subsection{V-ETS Performance Across Mixing Proportions and Data Quantity}

\begin{figure}[h]
  \centering
  \includegraphics[width=0.6\linewidth]{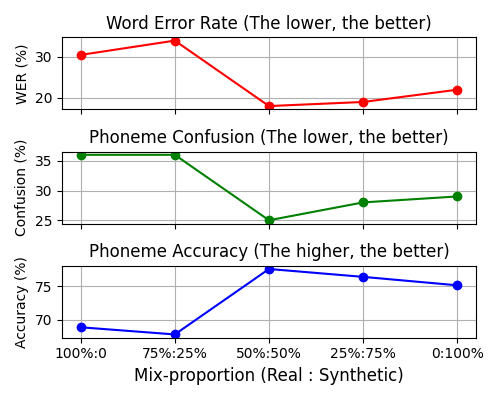}
  \caption{The evaluation results of WER, phoneme confusion and phoneme accuracy across different real-to-synthetic data ratios.}
  \label{fig:matrics_proportion}
\vspace{-0.2cm}
\end{figure}

To explore whether the effectiveness of self-training depends on the optimal mix proportion of real and synthetic data, we systematically test different training data mixing ratios to determine the optimal mix proportion of real and synthetic data. To ensure consistency, all models are trained using the same validation set drawn from real data (200 utterances). For evaluation, we construct a test set of 198 utterances, evenly split between 99 real and 99 synthetic utterances, allowing us to fairly compare the impact of different mix proportions.

The results are visualized in Figure \ref{fig:matrics_proportion}, where the x-axis represents different mix proportions: 100\%:0\%, 75\%:25\%, 50\%:50\%, 25\%:75\%, and 0\%:100\%, corresponding to the ratio of single-speaker real EMG data to multi-speaker synthetic Libri-EMG data. For evaluation, we employ three automatic metrics: WER, phoneme confusion rate, and phoneme accuracy. The results in Fig. \ref{fig:matrics_proportion} show that the model trained with a balanced mix of 50\% real and 50\% synthetic data outperforms all other configurations across all three evaluation metrics.

As mentioned in Sec. \ref{section:Mix-proportion}, we also evaluate the impact of dataset scaling on model performance under the optimal mix proportion 1:1. We trained our baseline ETS model on progressively larger datasets (1$\times$, 2$\times$, and 5$\times$1300 utterances) under the optimal 1:1 mix proportion. As shown in Table~\ref{tab:ThresholdWER}, increasing the dataset size consistently improved the performance in terms of all metrics:

\begin{table}[h]
    \centering
    \caption{Performance Comparison Across Dataset Sizes with real and synthetic data (Mix Proportion 1:1)}
    \begin{tabularx}{\linewidth}{|X|c|c|c|}
    \hline
    \makecell[l]{\textbf{Dataset Size}\\ \textbf{(50\% real + 50\% synthetic)}} & \textbf{WER} & \makecell{\textbf{Phoneme}\\\textbf{Confusion}} & \makecell{\textbf{Phoneme}\\\textbf{Accuracy}} \\ \hline
    1300 utt. ($\sim$3.2h) & 23.85 & 29.80 & 74.19 \\ \hline
    2$\times$1300 utt. ($\sim$6.4h) & 21.88 & 28.57 & 75.34 \\ \hline
    5$\times$1300 utt. ($\sim$16h) & \textbf{18.03} & \textbf{25.45} & \textbf{77.59} \\ \hline
    \end{tabularx}
    \label{tab:ThresholdWER}
\vspace{-0.3cm}
\end{table}

\begin{table*}[h!]
  \caption{The comparison of WER results across different Models and Datasets}
  \label{tab:scratch}
  \centering
  \begin{tabular}{p{4.2cm}|p{1.4cm}| p{3.1cm}|p{2.3cm}|p{3.8cm}}
    \toprule
     Test Dataset& \textbf{Baseline model \cite{gaddy2021improvedmodelvoicingsilent}}   & \textbf{Our CoM2S with self-training (mix ratio 1:1)}  & \textbf{Voiced baseline model \cite{gaddy2021improvedmodelvoicingsilent, Gaddy_EECS-2022-68}} & \textbf{Our CoM2S with mix-train-from-scratch (mix ratio 1:1)}           \\
        \midrule
    Real Single-speaker Data \cite{Gaddy_EECS-2022-68, gaddy2021improvedmodelvoicingsilent}                    & 	\multicolumn{1}{c|}{-} & 	\multicolumn{1}{c|}{-} & 	\multicolumn{1}{c|}{23.30\%\cite{Gaddy_EECS-2022-68}}&	\multicolumn{1}{c}{\textbf{21.87\%}}
         \\
    Our multi-speaker Libri-EMG  & 	\multicolumn{1}{c|}{54.21\%} & 	\multicolumn{1}{c|}{15.90\%} & 	\multicolumn{1}{c|}{37.63\%}& 	\multicolumn{1}{c}{\textbf{8.75\%}} \\

    \bottomrule
  \end{tabular}
\vspace{-0.5cm}
\end{table*}

As shown in Table \ref{tab:ThresholdWER}, increasing the dataset size from 3.2h to 16h of training data led to consistent improvements across all metrics: word error rate (WER) decreased by 24.4\% (from 23.85\% to 18.03\%), phoneme confusion reduced from 29.80\% to 25.45\%, and phoneme accuracy improved from 74.19\% to 77.59\%. This indicates that increasing the dataset size enhances model robustness and generalization.


\subsection{Scratch-Trained Model Evaluations}

As discussed in Sec. \ref{trainfromscratch}, we implement a controlled ablation study comparing two training paradigms: (1) a baseline model trained exclusively on real voiced EMG-speech pairs and (2) our proposed mixed-data model initialized with the previously determined optimal 1:1 real-synthetic ratio.

\subsubsection{Cross-Model Analysis with Baseline Data}

Table~\ref{tab:scratch} shows that our mix-train-from-scratch model achieves a WER of 21.87\% on the real single-speaker test set, outperforming the previous state-of-the-art WER of 23.30\% reported by the original voiced baseline~\cite{Gaddy_EECS-2022-68}. This suggests that synthetic pseudo-labeled EMG data can contribute positively to model learning, even when evaluated on real, natural articulatory inputs, validating both the effectiveness of our synthetic data generation and its benefits for representation learning through increased training diversity. 

\subsubsection{Cross-Dataset Generalization}
As supporting evidence, we also explored model generalization by evaluating all models on these two Libri-EMG datasets. As shown in Table~ \ref{tab:wer_comparison}, the baseline model performs poorly on this set (WER 54.21\%). In contrast, our mix-train-from-scratch model achieves a WER of 8.75\%, demonstrating strong generalization to speaker-independent data. Notably, even the self-training model initialized from the baseline (15.90\%) surpasses the baseline model by a large margin. 
These results indicate that the transduction model benefits from a more diverse training set, leading to better generalization across unseen data. 



\subsection{Subjective Evaluations by Human Listeners}
To complement our automatic metrics, we conducted a subjective evaluation study to assess our proposed CoM2S model in terms of speech intelligibility \cite{gaddy2021improvedmodelvoicingsilent} and speech quality \cite{ITU_P_800,hazan2013,ITU_R_BS} using two scratch-trained models, with both real and synthetic data as test sets. Two representative audio samples have been made available online \cite{chen2025vets_audio}.

\subsubsection{Speech Intelligibility}

Following a similar protocol to our automated transcription tests, we engaged one human evaluator who were unfamiliar with the target utterances. The evaluator listened to 20 randomly selected synthesized speech samples and transcribed what they perceived. 

\begin{table}[h]
    \centering
    \caption{Human WER Comparison Between Real and Synthetic Test Sets}
    \begin{tabular}{lcc}
        \hline
        Model & WER (real) & WER (synthetic)  \\
        \hline
        Voiced baseline model & 27.35\% & 27.04\% \\
        \makecell[l]{Our CoM2S with \\ mix-train-from-scratch} & \textbf{23.58\%} & \textbf{13.57\%} \\
        \hline
    \end{tabular}
    \label{tab:wer_comparison}
\vspace{-0.3cm}
\end{table}

As shown in Table~\ref{tab:wer_comparison}, the mix-train-from-scratch model achieves a 15.1\% relative reduction in WER on the real test set, confirming that synthetic data augmentation enhances generalization to real EMG-speech pairs. Notably, the model shows even stronger gains on synthetic test data (13.57\% WER, 49.8\% improvement over baseline), suggesting effective learning of synthetic patterns while maintaining real-world applicability. The persistent gap between real and synthetic performance highlights an opportunity to better align synthetic training data with real EMG characteristics in future work. 

\subsubsection{Speech Quality}
MOS (Mean Opinion Score) \cite{ITU_P_800, ITU_R_BS} is a subjective evaluation metric used to assess the perceived speech quality of synthesized or processed speech. Unlike objective metrics like WER, MOS captures human judgments of speech quality and overall listening experience. In our evaluation, ten evaluators rated the outputs on a 5-point scale (1: Bad, 5: Excellent) \cite{ITU_P_800}.

\begin{table}[h]
    \centering
    \caption{MOS Comparison Between Real and Synthetic Test Sets}
    \begin{tabular}{lcc}
        \hline
        Model & MOS (real) & MOS (synthetic)  \\
        \hline
        Voiced baseline model & 3.00 & 3.45 \\
        \makecell[l]{Our CoM2S with \\ mix-train-from-scratch} & \textbf{3.25} & \textbf{4.15} \\
        \hline
    \end{tabular}
    \label{tab:mos_comparison}
\vspace{-0.3cm}
\end{table}

The results are shown in Table~\ref{tab:mos_comparison}, the mix-train-from-scratch model achieves significantly higher MOS ratings than the real-only baseline on both real and synthetic test sets. Three key insights emerge: (1) The 8.3\% improvement on real data confirms that synthetic augmentation yields perceptibly higher-quality speech despite EMG artifacts; (2) The model’s superior performance on synthetic data (20.3\% higher MOS) suggests it successfully leverages multi-speaker diversity during training; (3) The baseline’s synthetic-set advantage (3.45 vs 3.00) implies inherent vocoder bias toward cleaner synthetic inputs. 
This improvement is notable given the frozen vocoder constraint, indicating that the gains stem primarily from the encoder’s improved EMG representation learning.

\section{Conclusions}
In this study, we investigate the use of synthetic EMG-speech data in self-training and enhance V-ETS model performance. 
Extensive experimental results confirm that our proposed CoM2S approach enhances phoneme recognition accuracy, reduces phonological confusion and word error rate, proving its effectiveness for V-ETS systems.
Subjective evaluations also verify the intelligibility of the generated speech from our proposed model by human listeners.
These results support the integration of synthetic data into future V-ETS training pipelines, potentially reducing reliance on large-scale real EMG recordings while maintaining high performance.
Building on our previous work \cite{Chen_Pitti_Quoy_Chen_2024}, we aim to extend the framework by introducing articulatory-level patterns derived from muscle activity in future studies.

\section*{Acknowledgment}
Xiaodan Chen is supported by CYU-IPAL and the A*STAR Research Attachment Program. 
Mathias Quoy was supported by CNRS funding for a research semester at the IPAL Lab in Singapore.  
This research is supported by A*STAR under its Japan-Singapore Joint Call: Japan Science and Technology Agency (JST) and A*STAR 2024 (R24I6IR136), and by the National Research Foundation, Singapore under its National Large Language Models Funding Initiative. Any opinions, findings, conclusions, or recommendations expressed in this material are those of the author(s) and do not reflect the views of the National Research Foundation, Singapore.


\bibliographystyle{IEEEtran}  
\bibliography{Bibliography}

@misc{chen2025vets_github,
  author       = {Xiaodan Chen},
  title        = {{The Dataset and Code Repository}},
  year         = {2025},
  howpublished = {\url{https://github.com/XiaodanChenSheldan/v-ets}},
  note         = {GitHub repository}
}

@dataset{chen2025vets_dataset,
  author       = {Xiaodan Chen},
  title        = {Libri-EMG Dataset},
  year         = {2025},
  version      = {1},
  publisher    = {Zenodo},
  doi          = {10.5281/zenodo.16788832},
  url          = {https://zenodo.org/records/16788832}
}

@misc{chen2025vets_audio,
  author = {Xiaodan Chen},
  title = {Audio samples},
  year = {2025},
  howpublished = {\url{https://xiaodanchensheldan.github.io/v-ets/}},
  note = {Accessed: 2025-08-10}
}

@article{Gao_Birkholz_Li_2024, title={Articulatory Copy Synthesis Based on the Speech Synthesizer VocalTractLab and Convolutional Recurrent Neural Networks}, volume={32}, url={http://dx.doi.org/10.1109/taslp.2024.3372874}, DOI={10.1109/taslp.2024.3372874}, journal={IEEE/ACM Transactions on Audio, Speech, and Language Processing}, publisher={Institute of Electrical and Electronics Engineers (IEEE)}, author={Gao, Yingming and Birkholz, Peter and Li, Ya}, year={2024}, pages={1845–1858} }

@misc{Chung_Kang_2024, title={Speaker-Independent Acoustic-to-Articulatory Inversion through Multi-Channel Attention Discriminator}, url={http://dx.doi.org/10.21437/Interspeech.2024-1269}, DOI={10.21437/interspeech.2024-1269}, journal={Interspeech 2024}, publisher={ISCA}, author={Chung, Woo-Jin and Kang, Hong-Goo}, year={2024}, month=sep, pages={1540–1544} }

@misc{lee2025articulatoryfeaturepredictionsurface,
      title={Articulatory Feature Prediction from Surface EMG during Speech Production}, 
      author={Jihwan Lee and Kevin Huang and Kleanthis Avramidis and Simon Pistrosch and Monica Gonzalez-Machorro and Yoonjeong Lee and Björn Schuller and Louis Goldstein and Shrikanth Narayanan},
      year={2025},
      eprint={2505.13814},
      archivePrefix={arXiv},
      primaryClass={eess.AS},
      url={https://arxiv.org/abs/2505.13814}, 
}

@article{Guenther_1995, title={Speech sound acquisition, coarticulation, and rate effects in a neural network model of speech production.}, volume={102}, url={http://dx.doi.org/10.1037/0033-295X.102.3.594}, DOI={10.1037/0033-295x.102.3.594}, number={3}, journal={Psychological Review}, publisher={American Psychological Association (APA)}, author={Guenther, Frank H.}, year={1995}, pages={594–621}, language={en} }

@article{Kearney_Guenther_2019, title={Articulating: the neural mechanisms of speech production}, volume={34}, url={http://dx.doi.org/10.1080/23273798.2019.1589541}, DOI={10.1080/23273798.2019.1589541}, number={9}, journal={Language, Cognition and Neuroscience}, publisher={Informa UK Limited}, author={Kearney, Elaine and Guenther, Frank H.}, year={2019}, month=mar, pages={1214–1229}, language={en} }

@article{Schultz_Wand_2010, title={Modeling coarticulation in EMG-based continuous speech recognition}, volume={52}, url={http://dx.doi.org/10.1016/j.specom.2009.12.002}, DOI={10.1016/j.specom.2009.12.002}, number={4}, journal={Speech Communication}, publisher={Elsevier BV}, author={Schultz, Tanja and Wand, Michael}, year={2010}, month=apr, pages={341–353}, language={en} }

@article{Silva_Moraes_Pernambuco_Moraes_Balata_2014, title={Use of surface electromyography in phonation studies: an integrative review}, volume={17}, url={http://dx.doi.org/10.7162/s1809-977720130003000014}, DOI={10.7162/s1809-977720130003000014}, number={03}, journal={International Archives of Otorhinolaryngology}, publisher={Georg Thieme Verlag KG}, author={Silva, Hilton and Moraes, Kyvia and Pernambuco, Leandro and Moraes, Sílvia and Balata, Patricia}, year={2014}, month=jan, pages={329–339}, language={en} }

@article{hazan2013,
  author = {Hazan, Valerie and Baker, Rachel},
  title = {How do we measure speech intelligibility?},
  journal = {Proceedings of Meetings on Acoustics},
  volume = {19},
  number = {1},
  pages = {060180},
  year = {2013},
  doi = {10.1121/1.4800694}
}

@misc{ITU_R_BS,
  author = {{International Telecommunication Union}},
  title = {ITU-R Recommendation {BS.1534-3}: Method for the subjective assessment of intermediate quality level of audio systems},
  howpublished = {ITU-R Recommendation BS.1534-3},
  month = {10},
  year = {2015},
  url = {https://www.itu.int/rec/R-REC-BS.1534},
  note = {Version 3},
  urldate = {2023-11-20}
}

@misc{ITU_P_800,
  author = {{International Telecommunication Union}},
  title = {Methods for subjective determination of transmission quality},
  howpublished = {ITU-T Recommendation P.800},
  year = {1996},
  note = {Version: 08/1996},
  url = {https://www.itu.int/rec/T-REC-P.800},
  urldate = {2023-09-15}
}

@misc{kong2020hifigangenerativeadversarialnetworks,
      title={HiFi-GAN: Generative Adversarial Networks for Efficient and High Fidelity Speech Synthesis}, 
      author={Jungil Kong and Jaehyeon Kim and Jaekyoung Bae},
      year={2020},
      eprint={2010.05646},
      archivePrefix={arXiv},
      primaryClass={cs.SD},
      url={https://arxiv.org/abs/2010.05646}, 
}

@article{DeepSpeech, title={Deep Speech: Scaling up end-to-end speech recognition}, url={https://arxiv.org/abs/1412.5567}, DOI={10.48550/ARXIV.1412.5567}, abstractNote={We present a state-of-the-art speech recognition system developed using end-to-end deep learning. Our architecture is significantly simpler than traditional speech systems, which rely on laboriously engineered processing pipelines; these traditional systems also tend to perform poorly when used in noisy environments. In contrast, our system does not need hand-designed components to model background noise, reverberation, or speaker variation, but instead directly learns a function that is robust to such effects. We do not need a phoneme dictionary, nor even the concept of a “phoneme.” Key to our approach is a well-optimized RNN training system that uses multiple GPUs, as well as a set of novel data synthesis techniques that allow us to efficiently obtain a large amount of varied data for training. Our system, called Deep Speech, outperforms previously published results on the widely studied Switchboard Hub5’00, achieving 16.0% error on the full test set. Deep Speech also handles challenging noisy environments better than widely used, state-of-the-art commercial speech systems.}, publisher={arXiv}, author={Hannun, Awni and Case, Carl and Casper, Jared and Catanzaro, Bryan and Diamos, Greg and Elsen, Erich and Prenger, Ryan and Satheesh, Sanjeev and Sengupta, Shubho and Coates, Adam and Ng, Andrew Y.}, year={2014} }

@techreport{mfa_english_us_arpa_acoustic_2024,
	author={McAuliffe, Michael and Sonderegger, Morgan},
	title={English (US) ARPA acoustic model v3.0.0},
	address={\url{https://mfa-models.readthedocs.io/acoustic/English/English (US) ARPA acoustic model v3\_0\_0.html}},
	year={2024},
	month={Feb},
}

@article{Morrison_Kumar_Kumar_Seetharaman_Courville_Bengio_2021, title={Chunked Autoregressive GAN for Conditional Waveform Synthesis}, url={https://arxiv.org/abs/2110.10139}, DOI={10.48550/ARXIV.2110.10139}, abstractNote={Conditional waveform synthesis models learn a distribution of audio waveforms given conditioning such as text, mel-spectrograms, or MIDI. These systems employ deep generative models that model the waveform via either sequential (autoregressive) or parallel (non-autoregressive) sampling. Generative adversarial networks (GANs) have become a common choice for non-autoregressive waveform synthesis. However, state-of-the-art GAN-based models produce artifacts when performing mel-spectrogram inversion. In this paper, we demonstrate that these artifacts correspond with an inability for the generator to learn accurate pitch and periodicity. We show that simple pitch and periodicity conditioning is insufficient for reducing this error relative to using autoregression. We discuss the inductive bias that autoregression provides for learning the relationship between instantaneous frequency and phase, and show that this inductive bias holds even when autoregressively sampling large chunks of the waveform during each forward pass. Relative to prior state-of-the-art GAN-based models, our proposed model, Chunked Autoregressive GAN (CARGAN) reduces pitch error by 40-60%, reduces training time by 58%, maintains a fast generation speed suitable for real-time or interactive applications, and maintains or improves subjective quality.}, publisher={arXiv}, author={Morrison, Max and Kumar, Rithesh and Kumar, Kundan and Seetharaman, Prem and Courville, Aaron and Bengio, Yoshua}, year={2021} }

@INPROCEEDINGS{Librispeech,
  author={Panayotov, Vassil and Chen, Guoguo and Povey, Daniel and Khudanpur, Sanjeev},
  booktitle={2015 IEEE International Conference on Acoustics, Speech and Signal Processing (ICASSP)}, 
  title={Librispeech: An ASR corpus based on public domain audio books}, 
  year={2015},
  volume={},
  number={},
  pages={5206-5210},
  doi={10.1109/ICASSP.2015.7178964}}

@article{Gao_Yue_Li_2022, title={Self-Transriber: Few-shot Lyrics Transcription with Self-training}, url={https://arxiv.org/abs/2211.10152}, DOI={10.48550/ARXIV.2211.10152}, abstractNote={The current lyrics transcription approaches heavily rely on supervised learning with labeled data, but such data are scarce and manual labeling of singing is expensive. How to benefit from unlabeled data and alleviate limited data problem have not been explored for lyrics transcription. We propose the first semi-supervised lyrics transcription paradigm, Self-Transcriber, by leveraging on unlabeled data using self-training with noisy student augmentation. We attempt to demonstrate the possibility of lyrics transcription with a few amount of labeled data. Self-Transcriber generates pseudo labels of the unlabeled singing using teacher model, and augments pseudo-labels to the labeled data for student model update with both self-training and supervised training losses. This work closes the gap between supervised and semi-supervised learning as well as opens doors for few-shot learning of lyrics transcription. Our experiments show that our approach using only 12.7 hours of labeled data achieves competitive performance compared with the supervised approaches trained on 149.1 hours of labeled data for lyrics transcription.}, publisher={arXiv}, author={Gao, Xiaoxue and Yue, Xianghu and Li, Haizhou}, year={2022} }

@article{Xu_Baevski_Likhomanenko_Tomasello_Conneau_Collobert_Synnaeve_Auli_2020, title={Self-training and Pre-training are Complementary for Speech Recognition}, url={https://arxiv.org/abs/2010.11430}, DOI={10.48550/ARXIV.2010.11430}, abstractNote={Self-training and unsupervised pre-training have emerged as effective approaches to improve speech recognition systems using unlabeled data. However, it is not clear whether they learn similar patterns or if they can be effectively combined. In this paper, we show that pseudo-labeling and pre-training with wav2vec 2.0 are complementary in a variety of labeled data setups. Using just 10 minutes of labeled data from Libri-light as well as 53k hours of unlabeled data from LibriVox achieves WERs of 3.0%/5.2% on the clean and other test sets of Librispeech - rivaling the best published systems trained on 960 hours of labeled data only a year ago. Training on all labeled data of Librispeech achieves WERs of 1.5%/3.1%.}, publisher={arXiv}, author={Xu, Qiantong and Baevski, Alexei and Likhomanenko, Tatiana and Tomasello, Paden and Conneau, Alexis and Collobert, Ronan and Synnaeve, Gabriel and Auli, Michael}, year={2020} }

@misc{Kahn_Lee_Hannun_2020, title={Self-Training for End-to-End Speech Recognition}, url={http://dx.doi.org/10.1109/ICASSP40776.2020.9054295}, DOI={10.1109/icassp40776.2020.9054295}, journal={ICASSP 2020 - 2020 IEEE International Conference on Acoustics, Speech and Signal Processing (ICASSP)}, publisher={IEEE}, author={Kahn, Jacob and Lee, Ann and Hannun, Awni}, year={2020}, month=may }

@misc{Wand_Janke_Schultz_2014, title={The EMG-UKA corpus for electromyographic speech processing}, url={http://dx.doi.org/10.21437/Interspeech.2014-379}, DOI={10.21437/interspeech.2014-379}, journal={Interspeech 2014}, publisher={ISCA}, author={Wand, Michael and Janke, Matthias and Schultz, Tanja}, year={2014}, month=sep }

@INPROCEEDINGS{7280404,
  author={Diener, Lorenz and Janke, Matthias and Schultz, Tanja},
  booktitle={2015 International Joint Conference on Neural Networks (IJCNN)}, 
  title={Direct conversion from facial myoelectric signals to speech using Deep Neural Networks}, 
  year={2015},
  volume={},
  number={},
  pages={1-7},
  doi={10.1109/IJCNN.2015.7280404}}

@INPROCEEDINGS{8578038,
  author={Diener, Lorenz and Felsch, Gerrit and Angrick, Miguel and Schultz, Tanja},
  booktitle={Speech Communication; 13th ITG-Symposium}, 
  title={Session-Independent Array-Based EMG-to-Speech Conversion using Convolutional Neural Networks}, 
  year={2018},
  volume={},
  number={},
  pages={1-5},
  doi={}}

@misc{Toth_Wand_Schultz_2009, title={Synthesizing speech from electromyography using voice transformation techniques}, url={http://dx.doi.org/10.21437/Interspeech.2009-229}, DOI={10.21437/interspeech.2009-229}, journal={Interspeech 2009}, publisher={ISCA}, author={Toth, Arthur R. and Wand, Michael and Schultz, Tanja}, year={2009}, month=sep, pages={652–655} }

@ARTICLE{8114359,
  author={Janke, Matthias and Diener, Lorenz},
  journal={IEEE/ACM Transactions on Audio, Speech, and Language Processing}, 
  title={EMG-to-Speech: Direct Generation of Speech From Facial Electromyographic Signals}, 
  year={2017},
  volume={25},
  number={12},
  pages={2375-2385},
  doi={10.1109/TASLP.2017.2738568}}

@misc{Diener_Vishkasougheh_Schultz_2020, title={CSL-EMG\_Array: An Open Access Corpus for EMG-to-Speech Conversion}, url={http://dx.doi.org/10.21437/Interspeech.2020-2859}, DOI={10.21437/interspeech.2020-2859}, journal={Interspeech 2020}, publisher={ISCA}, author={Diener, Lorenz and Vishkasougheh, Mehrdad Roustay and Schultz, Tanja}, year={2020}, month=oct, pages={3745–3749} }

@INPROCEEDINGS{10363027,
  author={Scheck, Kevin and Ivucic, Darius and Ren, Zhao and Schultz, Tanja},
  booktitle={Speech Communication; 15th ITG Conference}, 
  title={Stream-ETS: Low-latency End-to-end Speech Synthesis from Electromyography Signals}, 
  year={2023},
  volume={},
  number={},
  pages={200-204},
  doi={10.30420/456164039}}

@INPROCEEDINGS{10781707,
  author={Scheck, Kevin and Ren, Zhao and Dombeck, Tom and Sonnert, Jenny and van Gogh, Stefano and Hou, Qinhan and Wand, Michael and Schultz, Tanja},
  booktitle={2024 46th Annual International Conference of the IEEE Engineering in Medicine and Biology Society (EMBC)}, 
  title={Cross-Speaker Training and Adaptation for Electromyography-to-Speech Conversion}, 
  year={2024},
  volume={},
  number={},
  pages={1-4},
  doi={10.1109/EMBC53108.2024.10781707}}

@inproceedings{gaddy2021improvedmodelvoicingsilent,
    title = "An Improved Model for Voicing Silent Speech",
    author = "Gaddy, David and Klein,  Dan",
    editor = "Zong, Chengqing  and
      Xia, Fei  and
      Li, Wenjie  and
      Navigli, Roberto",
    booktitle = "Proceedings of the 59th Annual Meeting of the Association for Computational Linguistics and the 11th International Joint Conference on Natural Language Processing (Volume 2: Short Papers)",
    month = aug,
    year = "2021",
    address = "Online",
    publisher = "Association for Computational Linguistics",
    url = "https://aclanthology.org/2021.acl-short.23",
    doi = "10.18653/v1/2021.acl-short.23",
    pages = "175--181",
}

@misc{wu2024emgtospeechnecklaceformfactor,
      title={Towards EMG-to-Speech with a Necklace Form Factor}, 
      author={Peter Wu and Ryan Kaveh and Raghav Nautiyal and Christine Zhang and Albert Guo and Anvitha Kachinthaya and Tavish Mishra and Bohan Yu and Alan W Black and Rikky Muller and Gopala Krishna Anumanchipalli},
      year={2024},
      eprint={2407.21345},
      archivePrefix={arXiv},
      primaryClass={eess.AS},
      url={https://arxiv.org/abs/2407.21345}, 
}

@phdthesis{Gaddy_EECS-2022-68,
    Author= {Gaddy, David},
    Title= {Voicing Silent Speech},
    School= {EECS Department, University of California, Berkeley},
    Year= {2022},
    Month= {May},
    Url= {http://www2.eecs.berkeley.edu/Pubs/TechRpts/2022/EECS-2022-68.html},
    Number= {UCB/EECS-2022-68},
}

@misc{Chen_Pitti_Quoy_Chen_2024, title={Developmental Predictive Coding Model for Early Infancy Mono and Bilingual Vocal Continual Learning}, url={http://dx.doi.org/10.1007/978-3-031-72350-6\_2}, DOI={10.1007/978-3-031-72350-6\_2}, journal={Lecture Notes in Computer Science}, publisher={Springer Nature Switzerland}, author={Chen, Xiaodan and Pitti, Alexandre and Quoy, Mathias and Chen, Nancy F.}, year={2024}, pages={16–32}, language={en} }

@inproceedings{scheck23_interspeech,
  author={Kevin Scheck and Tanja Schultz},
  title={{STE-GAN: Speech-to-Electromyography Signal Conversion using Generative Adversarial Networks}},
  year=2023,
  booktitle={Proc. INTERSPEECH 2023},
  pages={1174--1178},
  doi={10.21437/Interspeech.2023-174},
  issn={2958-1796}
}

@inproceedings{van_Niekerk_2022,
   title={A Comparison of Discrete and Soft Speech Units for Improved Voice Conversion},
   url={http://dx.doi.org/10.1109/ICASSP43922.2022.9746484},
   DOI={10.1109/icassp43922.2022.9746484},
   booktitle={ICASSP 2022 - 2022 IEEE International Conference on Acoustics, Speech and Signal Processing (ICASSP)},
   publisher={IEEE},
   author={van Niekerk, Benjamin and Carbonneau, Marc-Andre and Zaidi, Julian and Baas, Matthew and Seute, Hugo and Kamper, Herman},
   year={2022},
   month=may }

@INPROCEEDINGS{Scheck2023SUE2S,
  author={Scheck, Kevin and Schultz, Tanja},
  booktitle={ICASSP 2023 - 2023 IEEE International Conference on Acoustics, Speech and Signal Processing (ICASSP)},
  title={Multi-Speaker Speech Synthesis from Electromyographic Signals by Soft Speech Unit Prediction},
  year={2023},
  pages={1-5},
  doi={10.1109/ICASSP49357.2023.10097120},
  url={https://www.csl.uni-bremen.de/cms/images/documents/publications/ScheckSchultz-ICASSP23.pdf},
}

@INPROCEEDINGS{ren2023selflearning,
  author={Zhao Ren and Kevin Scheck and Tanja Schultz},
  title={Self-learning and Active-learning for Electromyography-to-Speech Conversion},
  year={2023},
  booktitle={15th ITG Conference on Speech Communication},
  pages={1--5},
  note={to appear},
  doi={10.30420/456164048},
}

@article{ren2024diffets,
  title={Diff-ETS: Learning a Diffusion Probabilistic Model for Electromyography-to-Speech Conversion},
  author={Ren, Zhao and Scheck, Kevin and Hou, Qinhan and van Gogh, Stefano and Wand, Michael and Schultz, Tanja},
  journal={arXiv preprint arXiv:2405.08021},
  year={2024}
}

@inproceedings{gaddy-klein-2020-digital,
    title = "Digital Voicing of Silent Speech",
    author = "Gaddy, David  and
      Klein, Dan",
    editor = "Webber, Bonnie  and
      Cohn, Trevor  and
      He, Yulan  and
      Liu, Yang",
    booktitle = "Proceedings of the 2020 Conference on Empirical Methods in Natural Language Processing (EMNLP)",
    month = nov,
    year = "2020",
    address = "Online",
    publisher = "Association for Computational Linguistics",
    url = "https://aclanthology.org/2020.emnlp-main.445",
    doi = "10.18653/v1/2020.emnlp-main.445",
    pages = "5521--5530",
}

@misc{vaswani2017attention,
      title={Attention Is All You Need}, 
      author={Ashish Vaswani and Noam Shazeer and Niki Parmar and Jakob Uszkoreit and Llion Jones and Aidan N. Gomez and Lukasz Kaiser and Illia Polosukhin},
      year={2017},
      eprint={1706.03762},
      archivePrefix={arXiv},
      primaryClass={cs.CL}
}


\end{document}